\documentclass[twocolumn,nofootinbib,superscriptaddress]{revtex4-1}

\usepackage{slashed}

\usepackage[normalem]{ulem}

\newcommand{\rmi}{\mathrm{i}}

\usepackage{color}

\usepackage{epsfig}
\usepackage{epstopdf}
\usepackage{epsf}
\input{epsf.sty}
\usepackage{graphicx,amsmath,amssymb}
\usepackage{epsfig}
\allowdisplaybreaks

\def\ei{\end{itemize}}
\def\be{\begin{equation}}
\def\ee{\end{equation}}
\newcommand{\bea}{\begin{eqnarray}}
\newcommand{\eea}{\end{eqnarray}}

\usepackage{bbm,bm,amsmath,amssymb}

\usepackage{hyperref}

\def\K{{K\"{a}hler}}

\newcommand{\rf}[1]{(\ref{#1})}
\newcommand{\cN}{\mathcal{N}}

\def\K{K{\"a}hler}

\usepackage[usenames,dvipsnames,table]{xcolor}

\usepackage{float}

\usepackage{breakurl}

\begin{document}

 \title{\Large\rm {\bf   \boldmath  de Sitter Minima from M-theory and  String theory}}

\author{Niccol\`o Cribiori}  
\email{niccolo.cribiori@tuwien.ac.at}
\affiliation{Institute for Theoretical Physics, TU Wien,\\ Wiedner Hauptstrasse 8-10/136, A-1040 Vienna, Austria}
\author{Renata Kallosh}
\email{kallosh@stanford.edu}
\affiliation{Stanford Institute for Theoretical Physics and Department of Physics, Stanford University, Stanford,
CA 94305, USA}
\author{Andrei Linde}
\email{alinde@stanford.edu}
\affiliation{Stanford Institute for Theoretical Physics and Department of Physics, Stanford University, Stanford,
CA 94305, USA}
 \author{Christoph Roupec}
\email{christoph.roupec@tuwien.ac.at}
\affiliation{Institute for Theoretical Physics, TU Wien,\\ Wiedner Hauptstrasse 8-10/136, A-1040 Vienna, Austria}

 \begin{abstract}
 We study M-theory compactification on ${\mathbb{T}^7/ \mathbb{Z}_2^3}$ in  the presence of a seven-flux, metric fluxes  and KK monopoles. The effective four-dimensional supergravity has seven chiral multiplets whose couplings are specified by the $G_2$-structure of the internal manifold. We supplement the corresponding superpotential by a KKLT type  non-perturbative exponential contribution for all, or for some of the seven moduli, and find a discrete set of supersymmetric Minkowski minima. We also study type IIA and type IIB  string theory compactified on ${\mathbb{T}^6/ \mathbb{Z}_2^2}$. In type IIA, we use a six-flux, geometric fluxes and non-perturbative exponents. In type IIB theory, we use F and H fluxes, and non-geometric Q and P fluxes, corresponding to consistently gauged supergravity with certain embedding tensor components, \emph{without non-perturbative exponents}. Also in these situations, we produce discrete Minkowski minima. Finally, to construct dS vacua starting from these Minkowski progenitors, we follow the procedure of mass production of dS vacua.  \end{abstract}

\maketitle



\parskip 4 pt

\section{Introduction} 
In \cite{Kallosh:2019zgd,Cribiori:2019drf}, we introduced a method to construct de Sitter minima, starting from Minkowski minima in type IIA and type IIB string theory. Here, we  apply this method in the context of M-theory and string theory. All of our models here have seven complex scalars, which are coordinates of the coset space $\Big[{SL(2,\mathbb{R})\over SO(2)}\Big]^7$.

We begin with moduli stabilization in M-theory on a seven-manifold with $G_2$-structure, namely the \emph{twisted seven-torus}. The starting point is the compact manifold with $\mathbb{Z}_2\times \mathbb{Z}_2\times \mathbb{Z}_2\subset G_2$  holonomy that is obtained as the toroidal orbifold of the form $X_7={\mathbb{T}^7/  \mathbb{Z}_2\times \mathbb{Z}_2\times \mathbb{Z}_2}$, \cite{DallAgata:2005zlf,Duff:2010vy,Derendinger:2014wwa,Ferrara:2016fwe}. We make the quotient non-singular by a choice of a
free orbifold action\footnote{We are grateful to A. Braun for explaining this and related issues of $G_2$-structures to us.}. 
The  Betti numbers of $X_7$ are
$(b_0, b_1, b_2, b_3)= (1,0,0,7).$ 
This theory is identified with the maximal rank reduction on the  seven-torus  and leads directly to 4d $\cN=1$  supergravity with seven moduli. Then, the twisting is introduced and can be interpreted as a Scherk--Schwarz reduction on the original torus. To derive the  twisted seven-torus   model from M-theory, it was proposed in \cite{DallAgata:2005zlf} how  to generalize the action of  11d supergravity to its `democratic form',  namely a pseudo-action where the potentials and the dual curvatures appear at the same time. In 10d, this type of supergravity pseudo-action was proposed in \cite{Bergshoeff:2001pv}. The pseudo-action allows to identify the superpotentials in 4d supergravity, originating from M-theory on  twisted seven-tori. Following \cite{DallAgata:2005zlf,Derendinger:2014wwa}, below we discuss such superpotentials and use them to construct dS minima with all moduli stabilized. Another derivation of an effective 4d supergravity theory could also be done using the duality-symmetric 11d supergravity action coupled to M-branes \cite{Bandos:1997gd}.

M-theory on a  generalized twisted seven-torus  was proposed and  studied in \cite{Derendinger:2014wwa,Blaback:2018hdo}, following the corresponding beyond twisted tori constructions in 10d, given in \cite{Villadoro:2007yq}. In particular, the idea in \cite{Villadoro:2007yq} was to introduce Kaluza-Klein monopoles KK5 and KKO5-planes, which allow to consistently relax some restrictions, known as tadpole conditions. Then, in  \cite{Derendinger:2014wwa,Blaback:2018hdo}, an analogous construction was introduced and studied in M-theory. A  `beyond twisted tori'  construction was presented, by allowing the presence of KK6 monopoles and KKO6-planes.

The purpose of this note is to use M-theory on the generalized twisted seven-torus, to identify some relatively simple discrete supersymmetric Minkowski vacua, in which all of the 14 real scalars are stabilized. In turn, these vacua can be used to stabilize all of the 14 moduli in dS minima, following  the mechanism of mass production of dS vacua \cite{Kallosh:2019zgd,Cribiori:2019drf}.
This mechanism is applicable to any M-theory/string theory motivated superpotential satisfying certain conditions. However, all examples given in  \cite{Kallosh:2019zgd,Cribiori:2019drf} were based on the KL-type racetrack superpotentials containing at least two nonperturbative exponential terms for each of the moduli \cite{Kallosh:2004yh}.

In this paper we will show that, by taking into account  polynomial flux terms in superpotentials originating from M-theory/string theory, one can achieve dS vacuum stabilization in models with a single exponent for each field. Alternatively, by including additional flux contributions, we can stabilize dS vacua in models where only some of the moduli have exponential terms in the  superpotentials. Some of these M-theory models have also an interpretation as type IIA models compactified on ${\mathbb{T}^6/ \mathbb{Z}_2^2}$ with fluxes.

Finally, we will present a particular class of models in  type IIB string theory,  describing the seven moduli  compactified on ${\mathbb{T}^6/ \mathbb{Z}_2^2}$ with fluxes.  The origin of  one of the non-geometric fluxes in this model is subtle: it was conjectured in  \cite{Aldazabal:2006up} to be present, based on S-duality of the theory, once the geometric flux is introduced. We show that in this model one can construct stable dS vacua without using any non-perturbative exponential contribution in the superpotential. 

\section{Generalized twisted seven-torus }

Following the discussion in \cite{DallAgata:2005zlf,Derendinger:2014wwa}, where the seven-moduli model was derived from M-theory, we take the \K\ potential for the seven chiral superfields $\Phi^i$ to be\footnote{In \cite{Bobev:2019dik} the seven moduli model was derived from 11d supergravity compactified on $S^7$. It has the same \K\, potential as in \rf{7m} however, the superpotential is different, defined by the regular embedding of $[SU(1,1)]^7 $ into $E_{7(7)}$ and underlying octonian structure. The AdS critical points in these models were derived using Machine Learning software. We are grateful to N. Bobev, T. Fischbacher and K. Pilch for attracting our attention to these constructions.}
\begin{equation}
\label{7m}
K =  - \sum_{i=1}^7 \log\left( -\rmi (\Phi^i - \overline{\Phi}^i)\right)\,.
\end{equation}
The superpotential derived in \cite{DallAgata:2005zlf,Derendinger:2014wwa} has the generic form $W_{pert} = g_7 + G_i {\Phi}^i + {1\over 2} M_{ij} \Phi^i \Phi^j$. In the present work, we will use this superpotential, with two additional modifications. First, we set $G_i=0$, in order to have only constant and quadratic terms in the moduli. Second, we add to this superpotential a KKLT-type non-perturbative exponential term. Therefore, the resulting $W$ is
\be
W = g_7 + {1\over 2} M_{ij} \Phi^i \Phi^j + \sum_{i=1}^7A_{i} e^{\rmi a_i \Phi^i}\,.
\label{7mW}
\ee
Here, $g_7$ is a seven-flux contribution, whereas terms quadratic in the moduli originate from geometric fluxes.\footnote{ The seven-moduli case in \cite{Cribiori:2019drf} is now equivalent to M-theory on a \emph{seven-torus}  ${\mathbb{T}^7/ \mathbb{Z}_2^3}$, in the presence of a seven-flux. The terms quadratic in moduli, coming from twisting of the \emph{seven-torus} in M-theory (or, from geometric fluxes in IIA), were not used in \cite{Cribiori:2019drf}, but KL-type double exponents were added to a seven-flux instead. D6/O6 and anti-D6 are a reduction from M-theory to string theory of KK6(KKO6). The relation between D6 and KK monopole in 11d is known as oxidation, see \cite{Ortin:2015hya}.}
Here, all parameters in $W$ are real. The matrix $M_{ij}$ is symmetric and all of its diagonal elements vanish. Therefore, it has 21 parameters. 
One could generalize this setting and use racetrack superpotentials, following  \cite{Kallosh:2019zgd,Cribiori:2019drf}. In that case, dS vacuum stabilization is possible even in absence of the term $g_7 + {1\over 2} M_{ij} \Phi^i \Phi^j $. The goal of this paper is to explore alternative possibilities, using no more than a single nonperturbative exponential term  for each of the moduli.

The nonperturbative exponential terms might arise from wrapped M2-branes.  It was shown in \cite{Harvey:1999as} that in M-theory compactified on  manifolds of $G_2$  holonomy, membranes wrapped on  3-cycles induce nonzero corrections to the superpotential.   In the ${\mathbb{T}^7/ \mathbb{Z}_2^3}$ model there are seven  3-cycles. Therefore, one expects  exponents in $W$ for each of the seven moduli, where ${\rm Im}\,  \Phi_i$ are the volumes of these seven 3-cycles.
 
 To find supersymmetric Minkowski vacua, one has to solve the equations $\partial_i W=0$ and $W=0$. The first of these equations gives
\be
- \rmi a_{i} A_{i} e^{\rmi a_{i}\Phi_i} = M_{ij} \Phi^j  \,.
\ee
which can be solved for the coefficients $A_i$ of the non-perturbative terms, resulting in
\be
A_{i}= \rmi a_{i}^{-1}  e^{-\rmi a_{i}\Phi_i } M_{ij} \Phi^j\,.
\ee
We split $\Phi^i=  \theta^i +  i  \, \phi^i $ and note that the solution is consistent at $\theta^i=0$. Then, we substitute  the parameters $A_i$ evaluated at the extremum, $\phi^i= \phi^i_0 $,\, $\theta^i=\theta^i_0=0$, back into the superpotential. After that, we subtract from the expression of $W$ the constant term thus obtained. This allows us to fix the parameter $g_7$ and to satisfy also the equation $W=0$. This solves the problem of finding a supersymmetric Minkowski vacuum in the seven-moduli $\Phi^i$  model.

Therefore, given a free choice of parameters, following this path one can obtain a supersymmetric Minkowski state. We will often find that the number of free parameters is much greater than the number of equations, which may allow us to omit some of the terms in the superpotential and still obtain a supersymmetric Minkowski vacuum. However, if we want to implement the procedure proposed in  \cite{Kallosh:2019zgd} for producing dS minima, we have to require additionally that
the potential does not have flat directions, or, equivalently, that it has a positive definite mass matrix in the vacuum, corresponding to its second derivatives.
The mass matrix in a supersymmetric Minkowski vacuum is 
\begin{equation}
V_{i\bar\jmath}^{Mink} = m_{ik}g^{k\bar k}m_{\bar k \bar \jmath} = e^K W_{ik}g^{k\bar k}\bar W_{\bar k \bar \jmath}.
\end{equation}
Therefore, in the seven-moduli model we are considering, flat directions are given by the zero modes of 
\be
W_{ij} =\partial_i \partial_j W = M_{ij} -\delta_{ij} A_i a^2_i e^{{\rm i}a_i \Phi_i},
\ee
evaluated in the vacuum. Notice that, if we have exponents in all directions, as in \eqref{7mW}, the matrix $W_{ij}$ is a generic symmetric matrix, including non-vanishing diagonal terms. Since $g_{i\bar\jmath}$ is positive definite, one or more zero modes are in fact present in the mass matrix when
$
\det W_{ij} = 0.
$
However, as we will show in several examples, this is actually a quite restrictive condition, which does not hold in generic models, unless peculiar cancellations occur. Therefore, in general one expects that
\be
\det W_{ij} \neq 0
\ee
and no flat directions are present in the mass matrix. 

In our previous papers \cite{Kallosh:2019zgd,Cribiori:2019drf}, where only constant terms in $W$ were present, a KL-type double exponent was necessary for each direction in the moduli space, in order to obtain stable solutions.  All such models do not have flat directions, by construction.  
Meanwhile in the new set of models discussed in this paper one may encounter flat directions, but one can eliminate them by adding fluxes.  In each of the  models to be studied in this paper we found that the flat directions are absent in Minkowski vacua for a  broad range of parameters, i.e. no fine tuning is necessary.

Furthermore, by adding more flux contributions, one can eliminate some of the single exponents, and  by adding  extra contributions from S-dual fluxes, as in \rf{Andrei}, \emph{one can eliminate all of the exponents}, still without flat directions. This is one of the central, most unexpected results of this paper.

In the presentation of our examples, we split the seven-moduli in a type IIA  language, as 
\be
 \Phi^i= \{S,  \, T_I,   \, U_J\},  \qquad I, J=1,2,3.
\ee
For convenience, we keep the same notation also for type IIB examples in Section \ref{sec:IIBgauged}. Following also \cite{Derendinger:2014wwa}, the 21 non-vanishing terms contained in $M_{ij} $, in the case of effective supergravities coming from twisted reductions of M-theory on a $X_7=  {\mathbb{T}^7/ \mathbb{Z}_2^3}$ orbifold with fluxes, can be represented as:
\be
\begin{aligned}
\label{eq:effpara}
&{1\over 2} M_{ij} \Phi^i \Phi^j =  Sb^K U_K + U_I C^{IJ} T_J  \cr
& +  a^I \frac{U_1  U_2  U_3}{U_I} +  c^I \frac{T_1  T_2  T_3}{T_I} + S d^K T_K\,.
\end{aligned}
\ee
The 21 entries of $M_{ij}$ are now given in terms of the parameters $a^I,\, b^K,\, c^I,\, d^K$ and $C^{IJ}$. However, for our purpose of finding discrete  supersymmetric Minkowski vacua, it is sufficient to use  only some of these terms.

\vskip 5pt

\noindent {\it Model 1, with  S,  T and U exponents}

\noindent In this first class of models, we engage only 12 terms in $M_{ij}$ and keep one exponent for each of the seven directions. The resulting superpotential is then
\bea
&& W_1 =g_7 + b^K S U_K + C^{IJ} U_I T_J \nonumber \\\
\cr
&&+ A_{S} e^{\rmi a_S  S}+  \sum_IA_{T_I} e^{\rmi a_{T_I}  T_I} +\sum_IA_{U_I} e^{\rmi a_{U_I}  U_I}\,.
\label{bC}
\eea
In the vacuum, we have a total of 19 free parameters: 7  $a_i$,  3  $b^{I}$ and 9 parameters $C^{IJ}$. Instead of fixing all of  them and looking for the minimum of the potential, one can use 8 equations $\partial_i W=0$ and $W=0$ to find 8 parameters $g_7$ and $A_i$ such that  these equations  are satisfied at a chosen point $ \Phi^i$ in moduli space, which therefore describes a supersymmetric Minkowski vacuum. This still leaves plenty of free parameters to control the values of masses of all moduli in the vacuum and to ensure that there are no flat directions. As we will show in numerical examples, many options are available, even if one does not engage some of the exponents. We show one explicit example with unconstrained parameters, and another one where the tadpole conditions are satisfied without sources.

\noindent {\it Model 2, without S exponent}

A second class of models we consider is a subclass of the previous one, in which we set $A_S = a_S =0$ from the very beginning. In other words, we again use the 12 terms  from the $M_{ij}$ matrix and add the exponents in all of the directions but $S$. The superpotential in this case takes the form: 
\bea
W_2 = && g_7 +  b^K S U_K + C^{IJ} U_I T_J \cr
 \cr
&&+ \sum_I A_{T_I} e^{\rmi a_{T_I}  T_I} +\sum_I A_{U_I} e^{\rmi a_{U_I}  U_I}\,. 
\label{bCS}\eea
Solving $W=0$ and $\partial_i W=0$, in order to find a supersymmetric Minkowski solution, will now fix the parameters $g_7$, $A_{T_I}$, $A_{U_I}$, together with one of the parameters among $b^K$ or $C^{IJ}$, in \rf{bCS}. As it turns out, this does not prohibit a solution. Indeed, explicit examples of this class of models are possible.  We present  one such solution  in Section \ref{sec:bCexample}.

 \vskip 4.5pt

\noindent {\it Model 3, without U exponents}

\noindent In the third class of models, we consider 15 terms  from the matrix   $M_{ij}$ and add the exponents only in four directions, namely $S$ and $ T_I$. In particular,  it turns out that we do not need to add exponents in the $U_K$ directions. This is interesting, since in \cite{Cribiori:2019bfx,Cribiori:2019drf}  
such terms were employed in order to facilitate stable dS vacua in type IIA supergravity constructions, in which the only perturbative term in $W$ was a constant flux. Here we find that, by including in $W$ geometric fluxes polynomial in the moduli and looking first for a supersymmetric Minkowski minimum, some of the non-perturbative exponential  terms are not required. Therefore, we consider the superpotential
\bea
W_3 =&& g_7 + a^I \frac{U_1 U_2 U_3}{U_I} + b^K S U_K + C^{IJ} U_I T_J \cr
\cr
&&+ A_{S} e^{\rmi a_S  S} +  A_{T_I} e^{\rmi a_{T_I}  T_I}\, .
\label{abC}\eea
which has 24 parameters. Again, we have to solve the equations $\partial_i W=0$ and $W=0$, which will fix 8 parameters in $W$. We are then free to choose the remaining parameters in order to obtain appropriate masses in Minkowski. We present an explicit numerical example of this class of models  in section \ref{sec:bCexample}.

\vskip 5pt
\noindent{\it Model 4, without T and U exponents}

\noindent In this fourth class of models, we engage 18 terms  from the matrix $M_{ij}$ and add the exponential contribution only in one direction, namely $S$. In particular, in this case we find that there is no need to add exponents in the $T_I$ and $ U_K$ directions. Therefore, the superpotential is
\bea
W_4 = g_7 + && a^I \frac{U_1 U_2 U_3}{U_I} + b^K S U_K + C^{IJ} U_I T_J \cr
&&+   c^I \frac{T_1 T_2 T_3}{T_I} + A_{S} e^{\rmi a_S  S}\,,
\label{abCTT}\eea
which has 21 parameters. We solve the 8  equations $\partial_i W=0$ and $W=0$  once more and fix the remaining free parameters to produce Minkowski vacua without flat directions. We show a numerical realisation of this model in section \ref{sec:abCexample}.

\parskip 4.5pt

\section{Generalized twisted six-torus}
Following \cite{Villadoro:2007yq,Derendinger:2014wwa,Blaback:2018hdo}, in type IIA string theory compactified on ${ \mathbb{T}^6\over \mathbb{Z}_2 \times \mathbb{Z}_2}$ one finds that only 15 terms are available, out of the total 21 terms present in M-theory and given in \eqref{eq:effpara}.  In particular, the last two terms in \eqref{eq:effpara}, namely $  c^I \frac{T_1 T_2 T_3}{T_I} + S d^K T_K$, with 6 parameters, $c^I$ and $d^K$, are absent  in standard type IIA orientifold constructions. Furthermore, the six-flux $f_6$ in type IIA replaces the seven-flux $g_7$ of the M-theory models. In the notation of \cite{Blaback:2018hdo}, with $a = 1,2,3$ and  $m = 4,5,6$, the 3 terms $ a^I \frac{U_1 U_2 U_3}{U_I}$ correspond to two-fluxes $F_{am}$. The 3 terms of the form $b^ISU_I$ correspond to non-geometric fluxes, with $b^I$ defined by $ \omega_{mn}{}^c $. Finally, the 9 terms of the form $C^{IJ}U_{I} T_{J}$ correspond to  non-geometric fluxes, where $C^{IJ}$  is defined by $\omega_{bp}{}^m$, $\omega_{bc}{}^a $.
Thus, our {\it Models 1, 2 , 3} are also models in type IIA. Instead, our M-theory {\it Model 4} is not  related to standard type IIA orientifold constructions, due to the presence of the term $ c^I \frac{T_1 T_2 T_3}{T_I}$.

The   tadpole conditions require spacetime filling sources, such as O6 planes, D6 branes and KK monopoles, as explained in detail in  \cite{Villadoro:2007yq,Blaback:2018hdo}. In these cases, the combinations of fluxes $\sum_I a^I b^I $ and  $\sum_J a^J C^{JI}$ (see Table 2 in \cite{Blaback:2018hdo}) do not have to vanish, but can be canceled by specific O6/D6 sources. Similarly, the expressions  $b^{I}C^{IJ} + b^J C^{II}$ and  $C^{IJ} {C}^{JK} +  {C}^{IK}  {C}^{JJ}$  do not need to be set to zero, but can be cancelled by contributions from (KK5/KKO5) and from  (KK5/KKO5)' respectively,  where these sources are wrapped on specific  internal cycles.

In  {\it Models 3} and {\it 4}, we need to consider all of these conditions, while in  {\it Models 1} and {\it 2}  the first two are satisfied automatically, since $a^I=0$. The fact that the tadpole conditions can be satisfied in the presence of sources means, as it was already suggested in \cite{Villadoro:2007yq,Blaback:2018hdo},  that there is no need to enforce the Jacobi constraints on flux parameters, which would be required in absence of sources. Our examples will include one case where the tadpole identities are satisfied even without sources, as well as  more general cases with sources and relaxed Jacobi constraints.
\section{M-theory Examples}\label{sec:abCexample}
In this section, we investigate the models described above, in the context of an effective  4d $\mathcal{N} = 1$ supergravity  description and present numerical examples. The \K\, potential, in our conventions, takes the form \rf{7m} and the complete superpotential is given in eqs. \rf{7mW} and  \rf{eq:effpara}. After solving for the supersymmetric Minkowski vacuum and choosing the free parameters such that there are no flat directions, we follow the mass production mechanism \cite{Kallosh:2019zgd,Cribiori:2019drf}  in order to find a dS solution. We refrain from giving the details of this construction here and choose to present only the independent set of parameters and masses in Minkowski as well as in dS. 

One important comment concerns the uplifting procedure, which is well understood in both type IIB as well as type IIA string theory. It is based on pseudo-calibrated Dp-branes \cite{Kallosh:2018nrk} and results in an equivalent procedure of supplementing $4d$, $\mathcal{N} = 1$ supergravity by a nilpotent multiplet\footnote{Examples of dS vacua without an  uplifting  anti-Dp-brane (without a nilpotent multiplet)   based on perturbative and non-perturbative contributions are given in \cite{Blaback:2013qza} and based on higher derivative $R^4$ correction in M-theory models in  \cite{Blaback:2019zig}.}. In M-theory, the analogous procedure has not been worked out in detail yet and it will be a matter of future investigations. 

\vskip 2mm

\noindent {\it Model 1, with S,  T and U exponents}
\label{sec:bCexample}

\noindent The superpotential of this model is given in \eqref{bC}. For our first example, we  choose to solve the Minkowski conditions $W=0$ and $\partial_i W =0$ in terms of the parameters $A_S$, $A_{T_I}$, $A_{U_I}$ ($I=1,2,3$) and the seven-flux $g_7$. All of the other parameters, as well as the position of the minimum in moduli space, remain free. Then, we choose values for these free parameters in a way that avoids flat directions,   which might  happen in case of accidental cancellations, for very specific values of the parameters. One possible choice for the free parameters is given in Table \ref{tab:bC}. These parameters lead to a stable, supersymmetric Minkowski vacuum with canonical masses given in Table \ref{tab:bCmass}.
\begin{table}[htb]
\center
\begin{tabular}{|c|c||c|c||c|c||c|c|}\hline
 $\,S_0\,$     & $1.0$   & $\,a_S\,$ & $\;1.0\,$ &$\,C^{11}\,$ & $\,0.11\,$ &  $\,C^{32}\,$ & $\,0.32\,$\\\hline
 $T_{1,0}$ & $\,1.1\,$ & $\,a_{T_1}\,$ & $\;1.1\,$ & $C^{12}$ & $0.12$ & $C^{33}$ & $0.33$ \\\hline
 $T_{2,0}$ & $1.2$ & $\,a_{T_2}\,$ & $\;1.1\,$ & $C^{13}$ & $0.13$ & $b^1$     & $0.55$\\\hline
 $T_{3,0}$ & $1.3$ & $\,a_{T_3}\,$ & $\;1.1\,$ & $C^{21}$ & $0.21$ & $b^2$     & $0.60$  \\\hline
 $U_{1,0}$ & $5.1$ & $\,a_{U_1}\,$ & $\;0.51\,$ & $C^{22}$ & $0.22$ & $b^3$     & $0.65$ \\\hline
 $U_{2,0}$ & $5.2$ & $\,a_{U_2}\,$ & $\;0.52\,$ & $C^{23}$ & $0.23$ & $\Delta g_7$ & $5 \cdot 10^{-3} $ \\\hline
 $U_{3,0}$ & $5.3$ & $\,a_{U_3}\,$ & $\;0.53\,$ & $C^{31}$ & $0.31$ & $\mu^4$ & $ 9 \cdot 10^{-9} $ \\\hline
\end{tabular}
\caption{Our set of chosen parameters for {\it Model 1}. Note that $S_0$ corresponds to the imaginary part of the modulus, similarly for all of the other moduli. The values of the moduli $U_I$ are chosen in this way because, in our conventions, $Im(U_I)$ corresponds to the volume of the internal manifold, which should be large in IIA. Included are the downshift $\Delta g_7$ and uplift parameter $\mu^4$ for the mass production procedure.}
\label{tab:bC}
\end{table}

\begin{table}[htb]
\center
\resizebox{\linewidth}{!}{%
\begin{tabular}{|c|c|c|c|c|c|c|c|}\hline
     &$\,m_1\,$&$\,m_2\,$&$\,m_3\,$&$\,m_4\,$&$\,m_5\,$&$\,m_6\,$&$\,m_7\,$\\\hline
Mk & $\,0.6421\,$ & $\, 0.4700 \,$ & $\, 0.3216 \,$ & $\, 0.1757 \,$ & $\, 0.1406 \,$ & $\, 0.1129 \,$ & $\, 0.08219 \,$ \\\hline  
dS & $\, 0.6427 \,$ & $\, 0.4705 \,$ & $\, 0.3218 \,$ & $\, 0.1758 \,$ & $\, 0.1407 \,$ & $\, 0.1130 \,$ & $\,0.08227\,$ \\\hline  
\end{tabular}}
\caption{The canonical normalized masses for  {\it Model 1}. We choose to give only the masses of the moduli, omitting the axions. The behavior follows exactly as described in \cite{Cribiori:2019drf}.}
\label{tab:bCmass}
\end{table}
\noindent  {\it Model 1, with tadpole condition satisfied without sources}

\noindent Another interesting variation of the  model with $b^K $ and $C^{IJ}$ terms is connected to the tadpole conditions, as taken from Table 2 of \cite{Blaback:2018hdo}. Usually the tadpole conditions are satisfied by inclusion of sources. However, we find that, if we include exponents in all directions, we are able to satisfy all of the tadpole conditions without sources in this model. The relevant tadpole conditions, without sources, are:
\begin{equation}
\begin{aligned}
\label{eq:bCtad}
b^I C^{IJ} + b^J C^{II} &= 0, \\
C^{IJ} C^{JK} + C^{IK} C^{JJ}&=0\,,\qquad \text{(no summation).}
\end{aligned}
\end{equation}
We choose to solve these conditions  in terms of the $C^{IJ}$ with $I\neq J$,  keeping the other parameters as in Table \ref{tab:bC}. This leads to a stable solution with masses given in Table \ref{tab:massTad}.
\begin{table}[htb]
\center
\resizebox{\linewidth}{!}{%
\begin{tabular}{|c|c|c|c|c|c|c|c|}\hline
     &$\,m_1\,$&$\,m_2\,$&$\,m_3\,$&$\,m_4\,$&$\,m_5\,$&$\,m_6\,$&$\,m_7\,$\\\hline
Mk &  $\, 0.3006 \,$  &  $\, 0.1641 \,$  &  $\, 0.1179 \,$  & $\, 0.07467 \,$ &  $\, 0.06229 \,$  &  $\, 0.03988 \,$  &  $\, 0.02517 \,$  \\\hline  
dS &  $\, 0.2997 \,$  &  $\, 0.1637 \,$  &  $\, 0.1176 \,$  & $\, 0.07449 \,$  &  $\, 0.06227 \,$  &  $\, 0.03976 \,$  &  $\, 0.02513 \,$  \\\hline   
\end{tabular}}
\caption{The canonical normalized masses for {\it Model 1}  with all tadpole conditions solved.}
\label{tab:massTad}
\end{table}

\noindent {\it Model 2, without S exponent}

\noindent It is possible to set $A_S = 0$ from the very beginning, as given in \eqref{bCS}, in order to eliminate the non-perturbative contributions for the S-direction. Then, solving the 8 supersymmetric Minkowski equations for such a reduced model, gives a restriction on one of the flux parameters, for example $b^1$, besides the 7 parameters $g_7$, $A_{T_I}$ and $A_{U_I}$. Keeping all of the other parameters the same as in Table \ref{tab:bC}, leads to a stable solution, with masses given in Table \ref{tab:bCmassNoSexp}.
\begin{table}[htb]
\center
\resizebox{\linewidth}{!}{%
\begin{tabular}{|c|c|c|c|c|c|c|c|}\hline
     &$\,m_1\,$&$\,m_2\,$&$\,m_3\,$&$\,m_4\,$&$\,m_5\,$&$\,m_6\,$&$\,m_7\,$\\\hline
Mk & $\, 0.6360 \,$ & $\, 0.4629 \,$ & $\, 0.3295 \,$ & $\, 0.1491 \,$ & $\, 0.1225 \,$ & $\, 0.09989 \,$ & $\, 0.03602 \,$ \\\hline  
dS & $\, 0.6365 \,$ & $\, 0.4633 \,$ & $\, 0.3297 \,$ & $\, 0.1492 \,$ & $\, 0.1226 \,$ & $\, 0.09993 \,$ & $\, 0.03607 \,$ \\\hline  
\end{tabular}}
\caption{The canonical normalized masses of the moduli for the {\it Model 2}  without non-perturbative contributions for the $S$ direction. }
\label{tab:bCmassNoSexp}
\end{table}

\noindent {\it Model 3, without U exponents}

\noindent The superpotential of this model is given in \eqref{abC}. Compared to  {\it Model 1} and  {\it Model 2}, it contains an additional term $ a^I \frac{U_1 U_2 U_3}{U_I}$, which allows to build dS vacua without the U-exponent. When evaluating the conditions for supersymmetric Minkowski vacua, we can now solve for the three parameters $a^I$ (these $a^I$ parameters should not   be confused with the parameters in the exponents, $a_{\Phi_i}$).
 We find a stable dS solution with the same parameters as in Table \ref{tab:bC} and give the masses in Table \ref{tab:abCmass}.
\begin{table}[H]
\center
\resizebox{\linewidth}{!}{%
\begin{tabular}{|c|c|c|c|c|c|c|c|}\hline
     &$\,m_1\,$&$\,m_2\,$&$\,m_3\,$&$\,m_4\,$&$\,m_5\,$&$\,m_6\,$&$\,m_7\,$\\\hline
Mk & $\, 0.2569 \,$ & $\, 0.2342 \,$ & $\, 0.1706 \,$ & $\, 0.1424 \,$ & $\, 0.1260 \,$ & $\, 0.1030 \,$ & $\, 0.02566 \,$ \\\hline  
dS & $\, 0.2572 \,$ & $\, 0.2344 \,$ & $\, 0.1707 \,$ & $\, 0.1425 \,$ & $\, 0.1261 \,$ & $\, 0.1030 \,$ & $\, 0.02565 \,$  \\\hline  
\end{tabular}}
\caption{The canonical normalized masses for {\it Model 3}, without non-perturbative exponential corrections in the $U$-directions.}
\label{tab:abCmass}
\end{table}
 
\noindent {\it Model 4, without T and U exponents}

\noindent  The superpotential of {\it Model 4} is defined in \eqref{abCTT}. Including the terms $ c^I \frac{T_1 T_2 T_3}{T_I}$, from \eqref{abCTT}, we find that it is in fact possible to find a Minkowski solution without any exponents other than $A_S e^{\rmi a_s S}$, i.e. we set $A_{T_I} = A_{U_I} = 0$ for all $I$. Instead of solving for the pre-factors of the exponents in the $T$ and $U$ directions, we now obtain the solutions in terms of the parameters $a^I$ and $c^I$ of the terms quadratic in $U$- and $T$-moduli. Once again, we use the parameters of Table \ref{tab:bC} and obtain the Minkowski and dS masses for the moduli given in Table \ref{tab:onlySmass}. Once more, we found a stable dS solution after the mass production procedure.
\begin{table}[htb]
\center
\resizebox{\linewidth}{!}{%
\begin{tabular}{|c|c|c|c|c|c|c|c|}\hline
     &$\,m_1\,$&$\,m_2\,$&$\,m_3\,$&$\,m_4\,$&$\,m_5\,$&$\,m_6\,$&$\,m_7\,$\\\hline
Mk & $\, 0.2639 \,$ & $\, 0.2520 \,$ & $\, 0.1469 \,$ & $\, 0.06163 \,$ & $\, 0.04579 \,$ & $\, 0.03365 \,$ & $\, 0.02874 \,$ \\\hline  
dS & $\, 0.2636 \,$ & $\, 0.2513 \,$ & $\, 0.1467 \,$ & $\, 0.06163 \,$ & $\, 0.04565 \,$ & $\, 0.03363 \,$ & $\, 0.02871 \,$  \\\hline  
\end{tabular}}
\caption{The canonical normalized masses for the model with only one exponent, in the $S$-direction.}
\label{tab:onlySmass}
\end{table}

To summarize the results obtained so far,  in {\it Model 2}, {\it Model 3} and {\it Model 4} we find that  quadratic tree-level contributions to the superpotential can take the place of some of the the non-perturbative exponential terms that are usually required.

\section{IIB theory, gauged supergravity and dS vacua}\label{sec:IIBgauged}

In this section, we continue the investigation of the seven-moduli model with the \K\, potential given in \rf{7m}. The superpotential $W$ of the type IIB theory \cite{Aldazabal:2006up,Dibitetto:2011gm,Blaback:2013ht} has the following 3 structures: contributions coming from the $F$-flux, from the $H$-flux and from the $Q$-flux, which are all known fluxes in type IIB string theory. In addition, it was conjectured in  \cite{Aldazabal:2006up} that certain $P$-fluxes should be present due to S-duality of string theory. In \cite{Dibitetto:2011gm}, it was recognized that terms in $W$ of the form coming from the conjectured $P$-fluxes appear naturally as components of gauged supergravity in 4d, when the embedding tensor procedure is performed consistently.

For our purpose  we will keep only  terms  even in the moduli in the superpotential, namely we will use
 \bea\label{Andrei}
W_5&=&a_0 + a^{I} \frac{U_1 U_2 U_3}{U_I}\cr
\cr
& +&S\,\bigl({b^{I}\,U_{I}}\,+\,b_{3}\,U_{1}\,U_{2}\,U_{3}\bigr)\nonumber\\
&+& T_{K} \left(  C^{IK}\,U_{I} \, - \,c^{K}\,U_{1}\,U_{2}\,U_{3}\right)\cr &-& S\, T_{K} \left(d^{K} -  D^{IK}  \frac{U_1 U_2 U_3}{U_I}\right).
\eea
The first, second, third and fourth line represent the even parts of $F$-,$H$- ,$Q$- and $P$- flux, respectively. We find that the terms  with coefficients $b_3$ and $D^{IK}$  are not necessary for full stabilization of moduli, in this model. One can use \eqref{Andrei}, with or without terms proportional to $b_3$ and $D^{IK}$, as a new model  \emph{which does not have non-perturbative exponents in $W$}. As a numerical example, we will describe below a model with $D^{IK}=0$ but $b_3 \neq 0$ and find that there is a Minkowski minimum without flat directions. This means that we were able to employ this model in order to get a dS minimum, using the technology developed in \cite{Kallosh:2019zgd,Cribiori:2019drf}.

\vskip 2mm 

\noindent{\it Model 5, without any exponents}

\noindent In order to find an explicit example of a dS vacuum from the above model, we again have to solve the Minkowski conditions, $W=0$ and $\partial_i W = 0$ where $i= {S,\, T_I,\, U_I}$ with $I=1,2,3$. This will fix 8 of the parameters in \eqref{Andrei}. We choose, in this case, to solve for the following set:
$a_0,\, a^{I},\, b_3$ and $c^{K}$.
For the position in moduli space, the downshift to AdS, $\Delta a_0 = \Delta g_7$, and uplift to dS, we choose the same values as in Table \ref{tab:bC}. These values are supplemented by the ones in Table \ref{tab:noExp}.

\begin{table}[htb]
\center
\resizebox{\linewidth}{!}{%
\begin{tabular}{|r|r||r|r||r|r||r|r||r|r|}\hline
$\,b^{1}$ & $\, 0.55 \,$  & $\,C^{11}$ & $ -0.11 \,$ & $\,C^{21}$ & $\, 0.21 \,$ & $\,C^{31}$ & $\, 0.31 \,$ & $\,d^{1}$ & $\, 5.1 \,$ \\\hline
$\,b^{2}$ & $\, 0.60 \,$ & $\,C^{12}$ & $\, 0.12 \,$ & $\,C^{22}$ & $ -0.22 \,$ & $\,C^{32}$ & $\, 0.32 \,$ & $\,d^{2}$ & $ -5.2 \,$ \\\hline
$\,b^{3}$ & $\, 0.65 \,$ & $\,C^{13}$ & $\, 0.13 \,$ & $\,C^{23}$ & $\, 0.23 \,$ & $\,C^{33}$ & $ -0.33 \,$ & $\,d^{3}$ & $\, 5.3 \,$ \\\hline
\end{tabular}}
\caption{The independent parameters for our {\it Model 5}. These produce the values for the masses in Table \ref{tab:noExpmass}. No particular fine-tuning is necessary.}
\label{tab:noExp}
\end{table}
\noindent We found a stable Minkowski solution and then were able to follow the mass production procedure to obtain a dS vacuum with masses given in Table \ref{tab:noExpmass}. We also found that it is easy to change the parameters and still have dS minima, without particular fine-tuning. This model is very interesting since it has only polynomial terms in the superpotential.

\begin{table}[htb]
\center
\resizebox{\linewidth}{!}{%
\begin{tabular}{|c|c|c|c|c|c|c|c|}\hline
     &$\,m_1\,$&$\,m_2\,$&$\,m_3\,$&$\,m_4\,$&$\,m_5\,$&$\,m_6\,$&$\,m_7\,$\\\hline
Mk & $\, 0.5392 \,$ & $\, 0.4551 \,$ & $\, 0.1037 \,$ & $\, 0.06185 \,$ & $\, 0.05355 \,$ & $\, 0.02389 \,$ & $\, 0.01263 \,$ \\\hline  
dS & $\, 0.5391 \,$ & $\, 0.4552 \,$ & $\, 0.1036 \,$ & $\, 0.06183 \,$ & $\, 0.05357 \,$ & $\, 0.02381 \,$ & $\, 0.01260 \,$  \\\hline  
\end{tabular}}
\caption{For the IIB model without exponents, where all contributions come from tree-level fluxes, we find these canonical masses for the moduli.}
\label{tab:noExpmass}
\end{table} 
\noindent
\section{Discussion}
 M-theory is  supposed to unify all of the consistent versions of superstring theory. At low energies it should be approximated by 11d supergravity. 
Furthermore, it should also describe various extended objects, like M2 and M5 branes, KK6 monopoles and KKO6-planes, such that extended objects of string theory, like Dp-branes and Op-planes are included. The existence of such a theory was first conjectured by  Witten in 1995. Some early papers on M-theory   include
\cite{Witten:1995em,Duff:1996aw,Sen:1998kr}  and more information can be found in   the books \cite{Ortin:2015hya,Johnson:2003gi}. A particularly relevant description of M-theory and 4d gauged supergravity is given in \cite{DallAgata:2005zlf,Villadoro:2007yq,Derendinger:2014wwa,Dibitetto:2011gm}. We are using these models in our construction of 4d dS vacua. The main issue in studies of specific models of dS minima in 4d gauged supergravity  is their motivation from string theory or M-theory.

 Here we focused on a model where seven complex scalars are coordinates of the coset space $\Big[{SL(2,\mathbb{R}/ SO(2)}\Big]^7$. This model is available in M-theory and in type IIA and type IIB string theory. As a technical tool for constructing dS minima, we use the method of mass production  of dS vacua proposed in  \cite{Kallosh:2019zgd,Cribiori:2019drf},  based on the the possibility to make parametrically small deformations (downshift and uplift) of a supersymmetric Minkowski vacuum state, without flat directions. In all of the cases, the uplift is due to the existence of the pseudo-calibrated anti-Dp-branes in string theory, which in 4d supergravity is equivalent to the presence of a nilpotent chiral multiplet \cite{Kallosh:2018nrk}. In M-theory, the details of the uplifting procedure need to be investigated.  We presented several classes of models with stable dS vacua, with numerical examples in {\it Models 1-5}. In  these models a better understanding of the role of the geometric fluxes and tadpole conditions will be required, based on earlier studies of these issues in \cite{Grana:2006kf,Grana:2013ila,Plauschinn:2018wbo}.

 In all of the models which we studied in M-theory, namely {\it Models  1, 2, 3, 4}, we used a superpotential $W$ with polynomial terms in the moduli,  of degree 0 and  2,  and a single non-perturbative KKLT-type exponent for some of the moduli, as shown in \eqref{eq:effpara}. This is different from the case without terms quadratic in the moduli, where supersymmetric Minkowski vacua without flat directions are possible with KL-type double set of exponents in every moduli direction \cite{Kallosh:2019zgd,Cribiori:2019drf}.  After adding  quadratic terms, we found  supersymmetric Minkowski vacua without flat directions by  engaging a single non-perturbative exponent for each  of the 7 moduli, or only for 4 of them, or only for the $S$ field. In all of the models of this kind, namely {\it Models  1, 2, 3, 4}, we found  locally stable dS minima.

Perhaps the most surprising result is the model in section V, in type IIB string theory, which we call {\it Model 5}. Only terms which are even polynomials in moduli,  of degree 0, 2, 4, are present in \rf{Andrei}, and no non-perturbative exponents are required. 
 In a model of 4d supergravity associated with IIB string theory presented in section V, all of the terms in the \K\, and superpotential are  identified with type IIB string theory.  
 The only somewhat unusual term in \eqref{Andrei}  is
$ S\, T_{K} d_{0}^{(K)}$.  
It  was conjectured to be present in type IIB theory in  \cite{Aldazabal:2006up}, to support $S$-duality. It is interesting that this same term is also present in M-theory in \eqref{eq:effpara}, as well as in a consistent gauged supergravity in  \cite{Dibitetto:2011gm}. We have constructed supersymmetric Minkowski minima without flat directions, and the corresponding dS minima in this seven-moduli model.
 \section*{Acknowledgement}
We are grateful to I. Bena, J. Bl\r{a}b\"{a}ck, A. Braun, G. Dibitetto, M. Grana, L. Martucci, E. Plauschinn, T. Van Riet, N. Warner, T. Wrase, I. Zavala and other participants of the workshop `de Sitter constructions in String theory' at Saclay,  for valuable discussions.   RK and AL  are supported by SITP and by the US National Science Foundation Grant  PHY-1720397, and by the  Simons Foundation Origins of the Universe program (Modern Inflationary Cosmology collaboration),  and by the Simons Fellowship in Theoretical Physics.
NC and CR  are supported by an FWF grant with the number P 302.65.
 CR is grateful to  SITP for the hospitality while this work was performed and to the Austrian Marshall Plan Foundation for making his stay at SITP possible.
 
\bibliography{lindekalloshrefs}
\bibliographystyle{utphys}

\end{document}